\documentclass[reqno]{amsart} % 12pt, {article} %
\usepackage{amssymb}
\usepackage{amsmath}
\usepackage{amsfonts}

\theoremstyle{remark}

\def\stackreb#1#2{\ \mathrel{\mathop{#1}\limits_{#2}}}
\newcommand{\nc}{\newcommand}

\nc{\rnc}{\renewcommand} \nc{\beq}{\begin{equation}}
\nc{\eeq}{\end{equation}} \nc{\beqa}{\begin{eqnarray}}
\nc{\eeqa}{\end{eqnarray}} \nc{\T}{\mathbb{T}}

\def\rank{\operatorname{rank}}

\begin{document}

\begin{flushright} AEI-2011-019 \end{flushright}

\title[$4d/3d$ superconformal indices/partition functions]
{From $4d$ superconformal indices \\
to $3d$ partition functions}

\author{F.~A.~H. Dolan}
\address{DAMTP, Wilberforce Rd., Cambridge CB3 0WA, England, UK}

\author[V.~P.~Spiridonov]{V.~P.~Spiridonov}
\address{Bogoliubov Laboratory of Theoretical Physics,
JINR, Dubna, Moscow Region 141980, Russia}

\author{G.~S.~Vartanov}
\address{Max-Planck-Institut f\"ur Gravitationsphysik, Albert-Einstein-Institut
14476 Golm, Germany; e-mail address: vartanov@aei.mpg.de}

\begin{abstract}
An exact formula for partition functions in $3d$ field theories was
recently suggested by Jafferis, and Hama, Hosomichi, and Lee. These
functions are expressed in terms of specific $q$-hypergeometric
integrals whose key building block is the double sine function (or
the hyperbolic gamma function). Elliptic hypergeometric integrals,
discovered by the second author, define $4d$ superconformal indices.
Using their reduction to the hyperbolic level, we describe a general
scheme of reducing $4d$ superconformal indices to $3d$ partition
functions which imply an efficient way of getting $3d$
$\mathcal{N}=2$ supersymmetric dualities for both SYM and CS theories
from the ``parent" $4d$ $\mathcal{N}=1$ dualities for SYM
theories. As an example, we consider explicitly the duality pattern
for $3d$ $\mathcal{N}=2$ SYM and CS theories with $SP(2N)$ gauge
group with the antisymmetric tensor matter.
\end{abstract}

\maketitle

%%\tableofcontents

\noindent
{\bf Introduction}

\smallskip

Superconformal indices (SCIs) of supersymmetric Yang-Mills field theories in
four dimensions \cite{index} may be usefully identified with elliptic
hypergeometric integrals which were discovered  in \cite{S1}.  This identification
was made in \cite{DO} where,
following earlier work of Romelsberger  \cite{index}, exact matching between indices for
standard Seiberg dual theories was demonstrated for classical gauge groups, as well as
being extended to Kutasov-Schwimmer dualities for large $N$. This was the first instance
where matching between SCIs for dual theories was explicitly shown for finite rank
gauge groups, tests until then involving large $N$ expansions, in the context of the AdS/CFT
correspondence, for instance, in \cite{index}.
The relation between SCIs and the theory of elliptic hypergeometric functions
was systematically investigated in \cite{SV1,SV2}, where many new $\mathcal{N}=1$
dualities were discovered and a physical meaning of various mathematical properties of
the elliptic hypergeometric integrals was recognized (see also \cite{gadde,V}).
SCIs provide perhaps currently the most rigorous and powerful mathematical tool for
testing $4d$ supersymmetric dualities
whereby indices for theories with quite different matter content may be shown to
coincide due to non-trivial recently discovered special function identities.

SCIs of $3d$ field theories have an essentially more involved
form (due to monopole contributions, that do not analogously arise for $4d$ theories) --
see \cite{Bhattacharya:2008zy,Kim:2009wb,Imamura:2011su,N8,
Krattenthaler:2011da} and references therein.
In \cite{Krattenthaler:2011da} an attempt was made to
find a connection between $4d$ and $3d$ SCIs, but
no simple relation was found. In the present work we
concentrate on the partition functions (PFs) of $3d$ theories
\cite{Dolan:2008vc,Kapustin:2010xq,Kapustin,Drukker:2010nc,
Hosomichi:2010vh,Jafferis:2010un,Hama:2010av,Hama:2011,Jafferis:2011ns,Willett:2011gp}.
More precisely, we demonstrate that certain of these
partition functions as well as duality relations among
different theories can be obtained by a reduction of $4d$ SCIs and
corresponding duality relations.

The study of $3d$ partition functions using the localization
technique was initiated by Kapustin, Willett, and Yaakov
\cite{Kapustin:2010xq} inspired by \cite{Pestun}. In the work of
Jafferis \cite{Jafferis:2010un} and Hama, Hosomichi, and Lee
\cite{Hama:2010av} a general recipe for evaluating $3d$ PFs
was suggested. It was found that these functions are
expressed in terms of the $q$-hypergeometric integrals admitting the
$|q|=1$ regime \cite{Stokman,ds:unit} (which are referred also as
the hyperbolic $q$-hypergeometric integrals) and having equal
quasiperiods $\omega_1=\omega_2$. In \cite{Hama:2011} this result
was generalized to arbitrary values of the quasiperiods $\omega_1$
and $\omega_2$.

Physically, the reduction of $4d$ supersymmetric field theories on
$R^3\times S^1$ to $3d$ theories on $R^3$  was discussed in detail
by Seiberg and Witten \cite{SW3d}.
Here we observe that there is an explicit connection between SCIs of
$4d$ supersymmetric field theories on $S^3 \times S^1$, and the
PFs of $3d$ theories on the squashed three-sphere $S_b^3$,
that allows also for
a recipe for conjecturing $3d$ duals. Technically, this fact is realized
by the reduction of elliptic hypergeometric integrals \cite{S1} to hyperbolic
$q$-hypergeometric integrals, which was rigorously established by
Rains in \cite{rai:limits}. A detailed consideration of such
limiting cases was given by van de Bult \cite{BultPhD} (see also
\cite{Spiridonov:2010em}). This suggests SCIs of $4d$ theories may
be more fundamental objects than the PFs of $3d$
theories in that the properties of the latter are inherited from
elliptic hypergeometric integrals, which, as functions, are more
general, nevertheless having a simpler form. While the connection
between $4d$ and $3d$ dualities for supersymmetric field theories
was explored in \cite{Aharony:1997bx} and studied further in the
context of the  three-dimensional analog of Seiberg duality in
\cite{K,A}, here the connection between $4d$ SCIs and $3d$ partition
functions gives a different perspective with strong predictive
power.

In the following, we illustrate how a reduction in $4d$ SCIs lead to
formulae equivalent to $3d$ PFs by considering
particular examples.  The same reduction may be applied essentially to
other $4d$ SCIs and expressions for $3d$ PFs
recovered so that these examples suffice to show the general procedure.
Since $4d$ SCIs for dual theories are obtained from transformation
formulae for these elliptic hypergeometric integrals, the same reduction
applied to these integrals yields corresponding $3d$ partition
functions for dual theories, from which matter fields content
and their representations may be read off.
While this procedure is applied to a few examples here,
obviously if more examples were considered, due to the
profusion of transformation formulae available for elliptic
hypergeometric integrals, a whole plethora of new dualities
for $3d$ theories would potentially be implied.

\bigskip

\noindent
{\bf Reduction from $4d$ superconformal indices to $3d$ partition
functions}

\smallskip

Briefly reviewing the index, denote, as in \cite{DO}, $\mathcal{N}=1$ superconformal
$SU(2,2|1)$ group generators by $J_3,J_{\pm},
\overline{J}_3, \overline{J}_{\pm}$, (for the $4d$ Lorentz group
$SO(3,1)\sim SU(2)\times SU(2)$),
$Q_{\alpha},\overline{Q}_{\dot\alpha}$, $\alpha, \dot{\alpha} = 1,2$, $P_\mu$,
$\mu=1,\ldots,4,$ (for supertranslations), $H$ (for dilations), $K_\mu,$
$S_{\alpha},\overline{S}_{\dot\alpha}$ (for special superconformal transformations),
and $R$ (for the $R$-symmetry group $U(1)_R$) so that, for $Q=\overline{Q}_{1 }$ and
$Q^{\dag}=-{\overline S}_{1}$,
\begin{equation}
\{Q,Q^{\dag}\}= 2{\mathcal H},\quad
\mathcal{H}=H-2\overline{J}_3-3R/2.
\label{susy}\end{equation}
Then the superconformal index \cite{index} is constructed
in terms a matrix integral involving generators commuting with $Q$ as
\begin{eqnarray}\nonumber && \makebox[-2em]{}
I(p,q,f_k) = \int_{G} d \mu(g)\, \text{Tr} \Big( (-1)^{\mathcal F}
p^{\mathcal{R}/2+J_3}q^{\mathcal{R}/2-J_3}
\\ && \makebox[4em]{} \times
e^{\sum_{a} g_aG^a} e^{\sum_{k}f_kF^k}\Big),
\quad \mathcal{R}= R + 2 \overline{J}_3, \label{Ind}\end{eqnarray}
where $d \mu(g)$ is the invariant
measure of the gauge group, $\mathcal{F}$ is the fermion number operator,
and $g_a,\, f_k$ are the chemical potentials
associated with the gauge $G$ and flavor $F$ group
generators $G^a$ and $F^k$, respectively.
Only those states in the cohomology of $Q$ contribute and thus the
$\mathcal{H}$ eigenvalue dependence is trivial.
Here, as in \cite{DO}, $t=(pq)^{1\over 2}$ keeps track of the conformal
dimensions of operators and thus $-{1\over 2}\ln p q$ may be identified with the circumference of the thermal
circle, relevant in one limit below.  \eqref{Ind} has been computed for free field theory on $\Bbb{R}^4$  via group
characters \cite{DO}, following from \cite{DOO},  and, for more general
theories on $S^3\times S^1$,
using field theory arguments by Romelsberger \cite{index}.

We discuss now the reduction of $4d$ SCIs to PFs for $3d$ $\mathcal{N}=2$
supersymmetric theories (SYM or CS) for a particular pair of dual
$\mathcal{N}=1$ SYM theories for which the SCIs match.  This reduction
may of course be applied generally to the SCIs for Seiberg duals
considered in \cite{DO,SV2}, and would be expected to further
indicate/test $3d$ dualities.
The chosen theories are of particular interest because they are related to the Selberg
integral, an object of fundamental importance in various fields of mathematical
physics. The complete set of related dualities was found only using the SCI
technique in \cite{SV1}, and the novel $4d$ dualities discovered there
yield evidence for novel $3d$ dualities, via reduction to PFs.

For the theories of interest, the electric theory has the gauge group $G=SP(2N)$ and global symmetry group
$SU(8) \times U(1) \times U(1)_R.$ They contain a vector superfield $V$ in
the adjoint representation of $SP(2N)$ and matter fields (collected in
the table below) given by eight chiral multiplets $Q$ forming the
fundamental representation $f$ of $SP(2N)$ and an
antisymmetric $SP(2N)$-tensor field $X$.
\begin{center}
\begin{tabular}{|c|c|c|c|c|}
  \hline
    & $SP(2N)$ & $SU(8)$ & $U(1)$ & $U(1)_R$ \\  \hline
  $Q$ & $f$ & $f$ & $- \frac{N-1}{4}$ & $\frac{1}{2}$ \\
  $X$ & $T_A$ & 1 & 1 & $0$ \\
\hline
\end{tabular}
\end{center}
For $N=1$, the field $X$ is absent and the group $U(1)$ is decoupled.

The electric theory SCI is described by the following
elliptic hypergeometric integral \cite{SV1}:
\begin{eqnarray}\nonumber     && \makebox[-2em]{}
I_E = \frac{(p;p)_{\infty}^N (q;q)_{\infty}^N }{2^N N!}
\Gamma((pq)^s;p,q)^{N-1} \int_{{\mathbb T}^N}\prod_{1 \leq i < k
\leq N} \frac{\Gamma((pq)^s z_i^{\pm 1} z_k^{\pm 1};p,q)}
{\Gamma(z_i^{\pm 1} z_k^{\pm 1};p,q) }
    \\     &&
     \makebox[6em]{} \times
\prod_{j=1}^N \frac{\prod_{i=1}^8  \Gamma((pq)^{r_Q} y_i z_j^{\pm
1};p,q)} {\Gamma(z_j^{\pm 2};p,q)} \frac{d z_j}{2 \pi \textup{i} z_j},
\label{SP2N1}\end{eqnarray}
where $r_Q = (1 - (N-1)s)/4$, $s$ and $y_i$ are chemical potentials
for the groups $U(1)$ and $SU(8)$, respectively, with $\prod_{i=1}^8y_i=1$.
Denoting $t=(pq)^s$ and $t_i=(pq)^{r_Q} y_i, i=1,\ldots,8$,
we have the constraints $|t|, |t_i|<1$ and the balancing condition,
$$
t^{2N-2}\prod_{i=1}^8 t_i \ = \ (pq)^2.
$$
Here $\Gamma(z;p,q)$ is the elliptic gamma function,
$$
\Gamma(z;p,q) \ = \ \prod_{i,j=0}^\infty \frac{1-z^{-1} p^{i+1}
q^{j+1}}{1-zp^iq^j}, \qquad |p|,|q|<1,
$$
used together with the conventions
$\Gamma(a,b;p,q) \equiv \Gamma(a;p,q) \Gamma(b;p,q)$,
$\Gamma(az^{\pm1};p,q) \equiv \Gamma(az;p,q) \Gamma(az^{-1};p,q)$.  Function
\eqref{SP2N1} is a two-parameter generalization of the elliptic
Selberg integral of \cite{vDS}
and a multidimensional extension of the elliptic analogue of the
Gauss hypergeometric function of \cite{S1,AA}.

There are 72 dual theories associated with the orbit of $W(E_7)$-Weyl group
which are split in four different classes \cite{SV1}.
The first class of magnetic theories has the same gauge group and the flavor group
${F} \ = \ SU(4) \times SU(4) \times U(1)_B \times U(1).$
Its field content is described in the table below
\begin{center}
\begin{tabular}{|c|c|c|c|c|c|c|}
  \hline
                & $SP(2N)$            & $SU(4)$    & $SU(4)$    & $U(1)_B$ & $U(1)$ & $U(1)_R$
\\  \hline
  $q$             & $f$                & $f$            & 1            & $-1$   &   $-\frac{N-1}{4}$   &  $\frac{1}{2}$
  \\
 $\widetilde{q}$& $f$   & 1            &$f$&  $1$   &  $-\frac{N-1}{4}$    &  $\frac{1}{2}$
 \\
  $x$             & $T_A$              & 1            &    1         & 0    &  1    & 0
  \\
  $M_J  $             & 1         & $T_A$            &    1       & $2$     &  $\frac{2J-N+1}{2}$   & 1
  \\
  $\widetilde{M}_J $    & 1      & 1            &    $T_A$       & $-2$    &  $\frac{2J-N+1}{2}$    & 1     \\
\hline
\end{tabular}
\end{center}
where $J=0,\ldots,N-1$. Corresponding superconformal index has the form
\begin{eqnarray}  \label{MagnDual} \nonumber &&
    I_M =
\prod_{J=0}^{N-1} \prod_{1 \leq i < j \leq 4} \Gamma((pq)^{r_{M_J}}
y_i y_j;p,q) \prod_{5 \leq i < j \leq 8}
\Gamma((pq)^{r_{\widetilde{M}_J}} y_i y_j;p,q)
    \\  \nonumber   &&  \makebox[2em]{} \times
\Gamma((pq)^{s};p,q)^{N -1}\frac{(p;p)_{\infty}^{N}
(q;q)_{\infty}^{N}}{2^N N!} \int_{{\mathbb T}^N} \prod_{1 \leq i < j
\leq N} \frac{\Gamma((pq)^{s} z_i^{\pm 1} z_j^{\pm 1};p,q)} {
\Gamma(z_i^{\pm 1} z_j^{\pm 1};p,q) }
    \\    && \makebox[2em]{}   \times
\prod_{j=1}^N \frac{\prod_{i=1}^4 \Gamma((pq)^{r_q}v^{-2} y_i z^{\pm
1}_j;p,q) \prod_{i=5}^8 \Gamma((pq)^{r_{\widetilde q}}v^2 y_i z^{\pm
1}_j;p,q)} {\Gamma(z_j^{\pm 2};p,q)}\frac{d z_j}{2 \pi \textup{i} z_j},
\label{IM1}\end{eqnarray}
where $v=\sqrt[4]{y_1y_2y_3y_4}$ and
\begin{eqnarray*}
r_q = r_{\widetilde{q}} =
\frac{1- (N-1)s}{4} ,  \qquad
r_{M_J} = r_{\widetilde{M}_J} = \frac{1 - (N-1-2J)s}{2}.
\end{eqnarray*}
The equality of indices $I_E=I_M$ was established in \cite{SV1} as a
direct consequence of the integral identities proved
in \cite{AA} for $N=1$ and in \cite{Rains} for arbitrary $N$.

Before describing the reduction to
PFs for $3d$ theories, their construction is briefly indicated.
According to \cite{Kapustin:2010xq,Jafferis:2010un,Hama:2010av,Hama:2011},
the partition function for $3d$ $\mathcal{N}=2$ SYM  theories has the form
\beq \label{PF_def}
Z(\underline{f}) \ = \ \int_{-\textup{i}\infty}^{\textup{i}\infty}
 \prod_{j=1}^{\rank G}du_j\,  J(\underline{u})
Z^{vec}(\underline{u}) \prod_{I} Z_{\Phi_I}^{chir}(\underline{f},\underline{u}).
\eeq
Here $f_k$ are the chemical potentials for the flavor symmetry group $F$,
$u_j$-variables are associated with the Weyl weights for the Cartan
subalgebra of the gauge group $G$. For the CS theory one has
$J(\underline{u})=e^{-\pi \textup{i} k \sum_{j=1}^{\rank G} u_j^2}$,
where $k$ is the level of CS-term, and for SYM theories one has
$J(\underline{u}) = e^{2 \pi \textup{i} \lambda \sum_{j=1}^{\rank G} u_j}$,
where $\lambda$ is the Fayet-Illiopoulos term. The terms
$Z^{vec}(\underline{u})$ and $Z_{\Phi_I}^{chir}(\underline{f},\underline{u})$
 in (\ref{PF_def}) are coming from the vector superfield and the matter fields,
correspondingly, and they are expressed in terms of the hyperbolic gamma functions, see
below.

Now we describe the key degenerating limit from SCIs to obtain PFs.
 First, we parameterize variables as in \cite{rai:limits},
$$
t=e^{2 \pi \textup{i} v\tau}, \quad t_i = e^{2 \pi
\textup{i} v \mu_i},\quad i=1, \ldots, 8, \ \ \ p=e^{2 \pi \textup{i} v
\omega_1}, \ q=e^{2 \pi \textup{i} v \omega_2},
$$
with the balancing condition
\beq 2(N-1)\tau + \sum_{i=1}^{8} \mu_i  = 2 (\omega_1 + \omega_2),
\eeq
and then take the limit $v \rightarrow 0$.
It assumes $pq\to 1$, the limit of vanishing radius for $S^1$,
but it is not identical to that since, effectively, we are left with
the squashed three-sphere $S_b^3$ instead of $S^3$. Although one has
also $p/q\to 1$, there still remain contributions in PF coming from the
operator $(p/q)^{J_3}$ in the SCI, and the squashing parameter
$b^2=\omega_1/\omega_2$ survives this limit. This is most easily seen in
terms of matching with the results for such PFs obtained in \cite{Hama:2011}.
To obtain PFs on the usual round sphere $S^3$, one has to set $b=1$.

Using the notation of \cite{Spiridonov:2010em},
the limiting SCI takes the following form,
\begin{equation}
I_E\stackreb{=}{v\to 0}e^{-\pi \textup{i} (\omega_1 + \omega_2)
 ( 4 s N^2 - 6 (s-1) N + 2 s - 1 ) / 12 v \omega_1 \omega_2}I_E^r,
\label{lim_exp}\end{equation}
where
\begin{eqnarray}\label{Sp-el_red} &&
I^{r}_E(\mu_1, \ldots, \mu_8, \tau ; \omega_1,\omega_2) =   \frac{1}{2^N N!}
\gamma^{(2)}(\tau ;\omega_1,\omega_2)^{N-1}
\\ \nonumber && \makebox[-2.5em]{} \times \int_{-\textup{i} \infty}^{\textup{i} \infty} \prod_{1 \leq i <
k \leq N} \frac{\gamma^{(2)}(\tau \pm u_i \pm
u_k;\omega_1,\omega_2)}{\gamma^{(2)}(\pm u_i \pm
u_k;\omega_1,\omega_2)} \prod_{j=1}^N \frac{\prod_{i=1}^{8}
\gamma^{(2)}(\mu_i \pm u_j;\omega_1,\omega_2)}{\gamma^{(2)}(\pm 2
u_j;\omega_1,\omega_2)} \prod_{j=1}^{N} \frac{d u_j}{\textup{i}
\sqrt{\omega_1 \omega_2}}.
\end{eqnarray}
Here,
\beq \gamma^{(2)}(u;\omega_1,\omega_2) = e^{-\pi \textup{i}
B_{2,2}(u;\omega_1,\omega_2)/2} \frac{(e^{2 \pi i u/\omega_1}
\widetilde{q};\widetilde{q})_\infty}{(e^{2 \pi i
u/\omega_2};q)_\infty},\eeq
with the redefined base parameter,
\beqa\nonumber
 q = e^{2 \pi i \omega_1/\omega_2},\qquad  \widetilde{q} = e^{-2 \pi i
\omega_2/\omega_1},\eeqa and for $B_{2,2}(u;\omega_1,\omega_2)$
denoting the
second order Bernoulli polynomial, \beq B_{2,2}(u;\omega_1,\omega_2) =
\frac{u^2}{\omega_1\omega_2} - \frac{u}{\omega_1} -
\frac{u}{\omega_2} + \frac{\omega_1}{6\omega_2} +
\frac{\omega_2}{6\omega_1} + \frac 12.\eeq
The conventions,
$
\gamma^{(2)}(a,b;\omega_1,\omega_2) \equiv
\gamma^{(2)}(a;\omega_1,\omega_2) \gamma^{(2)}(b;\omega_1,\omega_2),
$
and
$
\gamma^{(2)}(a\pm u;\omega_1,\omega_2) \equiv
\gamma^{(2)}(a+u;\omega_1,\omega_2)
\gamma^{(2)}(a-u;\omega_1,\omega_2),
$
are applied throughout the paper.

A similar result can be obtained by considering the modified elliptic
hypergeometric integrals constructed from the modified elliptic gamma
function $G(u;\omega_1,\omega_2,\omega_3)$, \cite{AA}, after taking the
limit $\omega_3\to\infty$. In this case, no diverging factors emerge in
the reduction of integrals, see a detailed consideration of some
examples in \cite{ds:unit,Spiridonov:2010em}.
In Appendix A of \cite{Spiridonov:2010em}  different
forms of the function $\gamma^{(2)}(u)$ baring different names are listed.
In particular, $1/\gamma^{(2)}(u)$ is known as the double sine function.
In \cite{rai:limits,BultPhD} the  hyperbolic gamma
function $\Gamma_h(u)$ is used which is obtained after replacing
 in $\gamma^{(2)}(u)$ of $u$ by $\textup{i}u$ and the quasiperiods
$\omega_1,\omega_2$ by $\textup{i}\omega_1, \textup{i}\omega_2$ .

In the discussed limit, a computation of the asymptotics of $I_M$
yields the same diverging exponential as in \eqref{lim_exp}, and
one arrives at the relation $I_E^{r} = I_M^{r}$, where
\beqa\label{Vhyp}  &&
I^{r}_M(\mu_1, \ldots, \mu_8,\tau ; \omega_1,\omega_2) =
\frac{1}{2^N N!} \gamma^{(2)}(\tau ;\omega_1,\omega_2)^{N-1} \\ &&
\makebox[-2em]{} \times \prod_{j=0}^{N-1} \prod_{1 \leq i < k \leq
4} \gamma^{(2)}(j \tau + \mu_i + \mu_k;\omega_1,\omega_2) \prod_{5
\leq i < k \leq 8}
\gamma^{(2)}(j \tau + \mu_i + \mu_k;\omega_1,\omega_2) \nonumber \\
&& \makebox[-2em]{} \times \int_{-\textup{i} \infty}^{\textup{i}
\infty} \prod_{1 \leq i < k \leq N} \frac{\gamma^{(2)}(\tau \pm u_i
\pm u_k;\omega_1,\omega_2)}{\gamma^{(2)}(\pm u_i \pm
u_k;\omega_1,\omega_2)} \prod_{j=1}^N \frac{\prod_{i=1}^{8}
\gamma^{(2)}(\nu_i \pm u_j;\omega_1,\omega_2)}{\gamma^{(2)}(\pm 2
u_j;\omega_1,\omega_2)} \prod_{j=1}^{N} \frac{d u_j}{\textup{i}
\sqrt{\omega_1 \omega_2}},\nonumber\eeqa where
\begin{eqnarray} && \nu_j = \xi +
\mu_j, \ j=1,2,3,4, \qquad \nu_j = -\xi + \mu_j, \ j=5,6,7,8,
\nonumber \\ && \makebox[0em]{}
2 \xi = \omega_1 + \omega_2 - (N-1) \tau - \sum_{i=1}^4 \mu_i = -
\omega_1 - \omega_2 + (N-1) \tau + \sum_{i=5}^8 \mu_i.
\nonumber\end{eqnarray}
Applying the further limit,
 \beqa \label{lim} && \lim_{S \rightarrow \infty}
I^{r}_E(\mu_1,\ldots,\mu_6,\xi_1+\alpha S,\xi_2-\alpha S, \tau ; \omega_1,\omega_2)
\\ \nonumber && \makebox[2em]{} \times
e^{-\pi \textup{i} N ((\xi_2-\alpha S-\omega)^2 - (\xi_1+\alpha
S-\omega)^2)/\omega_1\omega_2},\eeqa where $\max \{\arg (\omega_1),
\arg (\omega_2)\}-\pi<\arg(\alpha)< \min \{\arg (\omega_1), \arg
(\omega_2)\}$ and $\omega =(\omega_1+\omega_2)/2$, with fixed $\mu_7
= \xi_1 + \alpha S$ and $\mu_8 = \xi_2 - \alpha S$, to \eqref{Vhyp},
carried out essentially in \cite{BultPhD}, leads to an expression
coinciding exactly with the partition function for $3d$
$\mathcal{N}=2$ SYM theory with $SP(2N)$ gauge group, six quarks and
one chiral field in absolutely antisymmetric representation of a
gauge group, namely,
 \beqa \label{awhyp} \nonumber
&& Z^{3d}_E = \frac{1}{2^N N!} \gamma^{(2)}(\tau
;\omega_1,\omega_2)^{N-1} \int_{-\textup{i} \infty}^{\textup{i}
\infty} \prod_{1 \leq i < k \leq N} \frac{\gamma^{(2)}(\tau \pm u_i
\pm u_k;\omega_1,\omega_2)}{\gamma^{(2)}(\pm u_i \pm
u_k;\omega_1,\omega_2)}
\\ && \makebox[3em]{} \times \prod_{j=1}^N \frac{\prod_{i=1}^{6}
\gamma^{(2)}(\mu_i \pm u_j;\omega_1,\omega_2)}{\gamma^{(2)}(\pm 2
u_j;\omega_1,\omega_2)} \prod_{j=1}^{N} \frac{d u_j}{\textup{i}
\sqrt{\omega_1 \omega_2}}. \label{hyp2}\eeqa
Note that only six variables $\mu_i, i=1,\ldots,6,$ and $\tau$ have survived and,
moreover, the balancing condition has disappeared.

The limit $S \rightarrow \infty$
is associated with the presence of an additional ``twisted instanton"
\cite{Aharony:1997bx,K,A}. In these papers the connection of $3d$ and
$4d$ dualities was discussed with the number of $3d$ flavors
less by one in comparison to the $4d$ case resembling our situation,
since we are ``integrating out" two quarks (one flavor) and get the
equality for PFs with six quarks (three flavors).

In \cite{BultPhD}, transformations of the integral in \eqref{hyp2}
forming the group $W(D_6)$ were deduced as a limit from
\eqref{Vhyp}. They lead to the representation,
\beqa
\label{dual_SPtr} &&
Z^{3d}_M = \frac{1}{2^N N!} \gamma^{(2)}(\tau
;\omega_1,\omega_2)^{N-1} \prod_{j=0}^{N-1} \prod_{1 \leq i < k \leq
4} \gamma^{(2)}(j \tau + \mu_i + \mu_k;\omega_1,\omega_2) \\
&& \makebox[3em]{} \times \prod_{j=0}^{N-1} \gamma^{(2)}(j \tau +
\mu_5 + \mu_6, 4 \omega - \sum_{i=1}^6 \mu_i -
(2N-j-2)\tau;\omega_1,\omega_2) \nonumber
\\ \nonumber && \makebox[-2em]{} \times \int_{-\textup{i} \infty}^{\textup{i} \infty} \prod_{1
\leq i < k \leq N} \frac{\gamma^{(2)}(\tau \pm u_i \pm
u_k;\omega_1,\omega_2)}{\gamma^{(2)}(\pm u_i \pm
u_k;\omega_1,\omega_2)} \prod_{j=1}^N \frac{\prod_{i=1}^{6}
\gamma^{(2)}(\nu_i \pm u_j;\omega_1,\omega_2)}{\gamma^{(2)}(\pm 2
u_j;\omega_1,\omega_2)} \prod_{j=1}^{N} \frac{d u_j}{\textup{i}
\sqrt{\omega_1 \omega_2}}, \eeqa
where the reflection identity,
$$\gamma^{(2)}(u,2\omega - u;\omega_1,\omega_2) = 1,$$
has been applied, and the transformed variables are given by,
\begin{eqnarray}\nonumber  &&
\nu_j = \xi + \mu_j, \quad j=1,2,3,4, \qquad \nu_j =
-\xi + \mu_j, \ j=5,6,
\\ \nonumber && \makebox[0em]{}
2 \xi =  \omega_1 + \omega_2  - (N-1) \tau - \sum_{i=1}^4 \mu_i.
\end{eqnarray}

Interpreting these integrals in terms of the $3d$ field theories,
we find the electric theory with the global symmetry group
$SU(6) \times U(1) \times U(1)_A \times U(1)_R$
and the field content as tabulated below,
\begin{center}
\begin{tabular}{|c|c|c|c|c|c|}
  \hline
   & $SP(2N)$ & $SU(6)$ & $U(1)$ & $U(1)_A$ & $U(1)_R$ \\
\hline
  $Q$ & $f$ & $f$ & $-\frac{N-1}{4}$ & 1 & $\frac 12$ \\
  $X$ & $T_A$ & $1$ & 1 & 0 & 0 \\
 \hline
\end{tabular}
\end{center}
These data may be directly read off from the partition
function (\ref{awhyp}). The denominator terms in the integral kernel
arise as the vector superfield contribution, while the numerator in the first
product of the kernel, as well as the pre-factor terms in front of the
integral, arise due the chiral superfield $X$, the
antisymmetric representation of $SP(2N)$, the rest comes from the
quarks in the fundamental representation. The parameters $\tau$ and
$\mu_i, i=1,\ldots,6$, absorbed already the hypercharges for the
abelian part of the global group $ \sum_{i=1}^3
r_i x_i,$ with $x_i,\, i=1,2,3,$ being the chemical potentials
of $U(1), U(1)_A$, and $U(1)_R$, respectively, and $r_i$  being
their hypercharges. More precisely, $x_1=\tau$, $x_2=\sum_{i=1}^6\mu_i/6$,
and $x_3=(\omega_1+\omega_2)/2$.

For the magnetic theory, the global symmetry group in the ultraviolet is
$SU(4) \times SU(2) \times U(1)_B \times U(1)\times U(1)_A \times U(1)_R,$ while
the matter content may be similarly tabulated as,
\begin{center}
\begin{tabular}{|c|c|c|c|c|c|c|c|}
  \hline
   & $SP(2N)$ & $SU(4)$ & $SU(2)$ & $U(1)_B$ & $U(1)$ & $U(1)_A$ & $U(1)_R$ \\
\hline
  $q_1$ & $f$ & $f$ & 1 & $-1$ & $-\frac{N-1}{4}$ & $-1$ & $\frac 12$ \\
  $q_2$ & $f$ & 1 & $f$ & 2 & $-\frac{N-1}{4}$ & $-1$ & $\frac 12$ \\
  $x$ & $T_A$ & 1 & 1 & 0 & 1 & 0 & 0 \\
  $M_{1,j}$ & $1$ & $T_A$ & 1 & 0 & $\frac{2j-N+1}{2}$ & 2 & 1 \\
  $M_{2,j}$ & $1$ & 1 & $T_A$ & 0 & $\frac{2j-N+1}{2}$ & 2 & 1 \\
  $Y_j$ & 1 & 1 & 1 & 0  & $\frac{2j-N+1}{2}$ & $-6$ & $1$ \\
 \hline
\end{tabular}
\end{center}
where $j=0,\ldots, N-1$.
Analogously to \cite{SV1}, the above mentioned transformations are supposed
to lead to $| W(D_6)/S_6| = 32$ different dual theories. We expect that integrals
\eqref{Sp-el_red} and \eqref{Vhyp} also have proper $3d$ interpretation,
which is a subject of separate consideration.

\bigskip

\noindent
{\bf $3d$ $\mathcal{N}=2$ SYM theory with $SP(2N)$ gauge group, $6f$ and
$T_A$}

\smallskip

As another example, here we discuss a $4d$
$s$-confining multiple duality considered in terms of indices in \cite{SV1}. The
flavor symmetry group is $F=SU(6) \times U(1)$ and the field content
of both theories is presented as follows,
\begin{center}
\begin{tabular}{|c|c|c|c|c|}
  \hline
    & $SP(2N)$ & $SU(6)$ & $U(1)$ & $U(1)_R$ \\  \hline
  $Q$ & $f$ & $f$ & $N-1$ & $\frac{1}{3}$ \\
  $A$ & $T_A$ & 1 & $-3$ & 0 \\
\hline \hline
  $A^k$ &   & 1 & $-3k$ & 0 \\
  $QA^mQ$ &   & $T_A$ & $2(N-1)-3m$ & $\frac 23$ \\  \hline
\end{tabular}
\end{center}
where $k=2,\ldots,N$ and $m=0,\ldots,N-1$.

The electric superconformal index is given by the elliptic Selberg
integral suggested by van Diejen and the second author in \cite{vDS},
\begin{eqnarray}\label{sp2_1}
&&I_E= \frac{(p;p)^{N}_{\infty} (q;q)^{N}_{\infty}}{2^NN!}
\Gamma(t;p,q)^{N-1} \int_{\mathbb{T}^N} \prod_{1 \leq i < k \leq N}
\frac{\Gamma(t z_i^{\pm 1} z_k^{\pm 1};p,q)}{\Gamma(z_i^{\pm 1}
z_k^{\pm 1};p,q)}
\\ \nonumber && \makebox[3em]{} \times \prod_{j=1}^{N}
\frac{\prod_{m=1}^{6} \Gamma(t_mz_j^{\pm 1};p,q)}{\Gamma(z_j^{\pm
2};p,q)}\prod_{j=1}^{N} \frac{d z_j}{2 \pi \textup{i} z_j}, \eeqa
while the magnetic index is,
\beqa\label{sp2_2} && I_M= \prod_{j=2}^N
\Gamma(t^j;p,q) \prod_{j=1}^N \prod_{1 \leq m < s \leq 6}
\Gamma(t^{j-1}t_mt_s;p,q),
\end{eqnarray}
where the balancing condition reads $t^{2N-2}\prod_{m=1}^6 t_m =pq.$

Integral (\ref{sp2_1}) can be reduced to the hyperbolic level in the
same way as before (see, e.g. \cite{BultPhD}). This yields (up to some
diverging factor which we skip for brevity) the
partition function of an electric $3d$ ${\mathcal N}=2$ SYM theory,
with $SP(2N)$ gauge group, four quarks and one chiral field in the
antisymmetric representation of the gauge group,  given by the
following formula,
\beqa \label{conf1} && Z^{3d}_E = \frac{1}{2^N
N!} \gamma^{(2)}(\tau;\omega_1,\omega_2)^{N-1} \int_{-\textup{i}
\infty}^{\textup{i} \infty} \prod_{1 \leq i < k \leq N}
\frac{\gamma^{(2)}(\tau \pm u_i \pm
u_k;\omega_1,\omega_2)}{\gamma^{(2)}(\pm u_i \pm
u_k;\omega_1,\omega_2)} \nonumber
\\ && \makebox[3em]{} \times \prod_{j=1}^N \frac{\prod_{i=1}^{4}
\gamma^{(2)}(\mu_i \pm u_j;\omega_1,\omega_2)}{\gamma^{(2)}(\pm 2
u_j;\omega_1,\omega_2)} \prod_{j=1}^{N} \frac{d u_j}{\textup{i}
\sqrt{\omega_1 \omega_2}}.\eeqa

Evidently, $Z^{3d}_E$ for this case can be evaluated exactly,
which can be seen by the direct consideration of the limit $v\to 0$ in
the relation $I_E=I_M$, yielding,
\beq
Z^{3d}_M \
= \  \prod_{j=2}^{N} \gamma^{(2)}(j \tau;\omega_1,\omega_2) \prod_{j=0}^{N-1} \frac{\prod_{1 \leq i < k \leq 4}
\gamma^{(2)}(j \tau + \mu_i + \mu_k;\omega_1,\omega_2)}{\gamma^{(2)}((2N-2-j)\tau
+ \sum_{i=1}^4 \mu_i;\omega_1,\omega_2)}.
\label{conf2} \eeq
The equality $Z^{3d}_E=Z^{3d}_M$ was
rigorously established for the first time by a different method in
\cite{ds:unit}, which result
was used as a motivation for a systematic consideration of the
reduction procedure in \cite{rai:limits}.

The dual theories obtained from the equality of the partition
functions both have the global symmetry group
$SU(4) \times U(1) \times U(1)_A \times U(1)_R.$  The spectrum of
the electric theory may be tabulated as follows,
\begin{center}
\begin{tabular}{|c|c|c|c|c|c|}
  \hline
   & $SP(2N)$ & $SU(4)$ & $U(1)$ & $U(1)_A$ & $U(1)_R$ \\
\hline
  $Q$ & $f$ & $f$ & 0 & 1 & $\frac 13$ \\
  $X$ & $T_A$ & $1$ & 1 & 0 & 0 \\
 \hline
\end{tabular}
\end{center}
while that for the confining magnetic theory may be similarly tabulated as,
\begin{center}
\begin{tabular}{|c|c|c|c|c|}
  \hline
   & $SU(4)$ & $U(1)$ & $U(1)_A$ & $U(1)_R$ \\
\hline
  $M_j=X^jQ^2$ & $T_A$ & $j$ & 2 & $\frac 23$ \\
  $N_k=X^k$ & $1$ & $k$ & 0 & 0 \\
  $Y_j$ & 1 & $-(2N-2-j)$ & $-4$ & $\frac 23$ \\
 \hline
\end{tabular}
\end{center}
where $k=2, \ldots, N$ and $j=0, \ldots, N-1$.

\bigskip

\noindent
{\bf Further dualities}

\smallskip

Further dualities are implied by subsequent reduction of the PFs implemented
by taking similar limits as in
(\ref{lim}). In contrast to the four-dimensional case where similar
reduction of SCIs corresponds to theories with fewer
flavors, here such reduction corresponds to theories that, whilst also having fewer
flavors than originally, have increased CS level. The technical details of the
reduction of corresponding integrals are skipped here and only the final
results for implied particular
dualities, without an exhaustive list, are indicated.

One such reduction describes a $3d$
$\mathcal{N}=2$ CS theory with gauge group $SP(2N)_{k/2}$ and
$SU(4-k) \times U(1) \times U(1)_A \times U(1)_R$ global symmetry
group with the spectrum involving, apart from the vector multiplet,
$4-k$ quarks $Q_i,i=1,\ldots,4-k,$ and one chiral superfield $X$ in
the antisymmetric representation of the gauge group. This is a
confining theory where the spectrum of the dual theory can be
directly read from the expressions of the corresponding partition
functions, following from the properties of the integrals presented
in Sect. 5.6.3 of \cite{BultPhD}. It involves singlets of the
$SU(4-k)$ flavor group $Y_j=X^{j+1},\, j=1, \ldots, N-1,$ and
baryons $M_j=X^jQ^2, j=1, \ldots, N$, described by the chiral
superfield in the antisymmetric representation of $SU(4-k)$,
see the table below,
\begin{center}
\begin{tabular}{|c|c|c|c|c|c|}
  \hline
   & $SP(2N)_{k/2}$ & $SU(4-k)$ & $U(1)$ & $U(1)_A$ & $U(1)_R$ \\
\hline
  $Q$ & $f$ & $f$ & 0 & 1 & $\frac 13$ \\
  $X$ & $T_A$ & $1$ & 1 & 0 & 0 \\
 \hline \hline
  $M_j=X^jQ^2$ & & $T_A$ & $j$ & 2 & $\frac 23$ \\
  $Y_l=X^l$ & & $1$ & $l$ & 0 & 0 \\
 \hline
\end{tabular}
\end{center}
where $l=2, \ldots, N$ and $j=0, \ldots, N-1$ and $k=1,\ldots,4$.
Note that for $k=3,4$ in the dual phase there are no fields $M_j$,
for $k=3$ there is only one quark in the electric theory, and for $k=4$ the electric
theory is just the CS theory with only one field $X$.

A particular example emerges from the above duality when $N=1$ and
$k=3$ which gives a $\mathcal{N}=2$ CS theory with $SP(2)_{3/2}$
gauge group and one quark. In the dual theory we have only the
contribution coming from additional topological sector as suggested
in \cite{Jafferis:2011ns} (where actually one has also some matter
field on the magnetic side since the authors consider a different
electric theory), reflected by the magnetic partition function
involving only an exponent with some phase.

Continuing the reduction of (\ref{conf1}) and (\ref{conf2}) in a
similar fashion, another limit can be identified with the confining
phase description of $3d$ $\mathcal{N}=2$ SYM theory with $U(N)$
gauge group and the global symmetry and matter fields as described in the
table below
\begin{center}
\begin{tabular}{|c|c|c|c|c|c|c|c|}
  \hline
   & $U(N)$ & $U(1)_l$ & $U(1)_r$ & $U(1)$ & $U(1)_A$ & $U(1)_J$ & $U(1)_R$ \\
\hline
  $Q$ & $f$ & 1 & 0 & 0 & 1 & 0 & $\frac 13$ \\
  $\widetilde{Q}$ & $f$ & 0 & 1 & 0 & 1 & 0 & $\frac 13$ \\
  $X$ & $T_A$ & 0 & 0 & 1 & 0 & 0 & 0\\
 \hline \hline
  $M_j=X^jQ\widetilde{Q}$ & & 1 & 1 & $j$ & 2 & 0 & $\frac 23$ \\
  $Y_l=X^l$ & & 0 & 0 & $j$ & 0 & 0 & 0 \\
  $W_{j,\pm}$ & & $-\frac 12$ & $-\frac 12$ & $-j$ & 0 & $\pm 1$ & $\frac 23$ \\
 \hline
\end{tabular}
\end{center}
where $j=0,\ldots,N-1$ and $l=2,\ldots,N$.
The exact evaluation
of the corresponding partition functions follows from Theorem 5.6.8
in \cite{BultPhD} (this theorem implies also dualities for CS theories
with $U(N)_{k/2}$ gauge group and one chiral field in the adjoint
representation and some number of flavors).

A whole set of dualities follows from appropriate reduction of
the PFs of $\mathcal{N}=2$ SYM theory with $SP(2N)$
gauge group, six flavors and one antisymmetric representation matter
field, (\ref{awhyp}) and (\ref{dual_SPtr}). These reductions
are related to the integrals described in Fig. 5.8
of \cite{BultPhD}. Corresponding models represent both SYM and CS
theories with different number of flavors and different CS level.
Specifically, the lines going to the left in Fig. 5.8 of
\cite{BultPhD} correspond in the field theory language to integrating
out matter fields, reducing the number of flavors by $1$ and
increasing the CS-level by $1/2$ each time, and the lines going to
the right correspond to a passage to $U(N)$ SYM or CS field theories
with adjoint matter, see \cite{Niarchos:2008jb} for similar
dualities involving two adjoint matter fields.

For example, one particular reduction of (\ref{awhyp}) and
(\ref{dual_SPtr}) leads to PFs for $\mathcal{N}=2$
CS theory with $SP(2N)_{1/2}$ gauge group, whose global symmetry
and the matter fields are described in the table below
\begin{center}
\begin{tabular}{|c|c|c|c|c|c|}
  \hline
   & $SP(2N)_{1/2}$ & $SU(5)$ & $U(1)$ & $U(1)_A$ & $U(1)_R$ \\
\hline
  $Q$ & $f$ & $f$ & $-\frac{N-1}{4}$ & 1 & $\frac 12$ \\
  $X$ & $T_A$ & $1$ & 1 & 0 & 0 \\
 \hline
\end{tabular}
\end{center}
This is a self-dual gauge group theory obeying the multiple
duality phenomenon arising from the $W(A_5)$ symmetry of the
partition function.
After application of a corresponding transformation formula,
the PF for an $\mathcal{N}=2$ CS theory arises whose global
symmetry group $SU(4) \times U(1)_f \times U(1) \times U(1)_A \times
U(1)_R $ differs from the original one (in a sense, this corresponds to the split of
the original $SU(5)$ group to $SU(4) \times U(1)$). The dual matter fields are
described in the following table
\begin{center}
\begin{tabular}{|c|c|c|c|c|c|c|}
  \hline
   & $SP(2N)_{1/2}$ & $SU(4)$ & $U(1)_B$ & $U(1)$ & $U(1)_A$ & $U(1)_R$ \\
\hline
  $q_1$ & $f$ & $f$ & $-1$ & $-\frac{N-1}{4}$ & $-1$ & $\frac 12$ \\
  $q_2$ & $f$ & 1 & 4 & $-\frac{N-1}{4}$ & $-1$ & $\frac 12$ \\
  $x$ & $T_A$ & 1 & 0 & 1 & 0 & 0 \\
  $M_{1,j}$ & $1$ & $T_A$ & 0 & $\frac{2j-N+1}{2}$ & 2 & 1 \\
 \hline
\end{tabular}
\end{center}
where $j=0,\ldots, N-1$.

A series of dualities stems from the reduction of PFs of
$\mathcal{N}=2$ SYM or CS theories with $SP(2N)$ gauge group, $N_f$
flavors and different CS levels using the results of
Sect. 5.5 of \cite{BultPhD}. These are the generalizations of the
Giveon-Kutasov \cite{Giveon:2008zn} type of dualities for $SP(2N)$
theories, see, e.g., \cite{Willett:2011gp} for a duality for
$SP(2N)_{k/2}$ CS theory with $N_f$ flavors.

\bigskip

\noindent
{\bf Conclusion}

\smallskip

This paper demonstrates that there is a deep relation between
superconformal indices for four-dimensional
field theories and partition functions of
three-dimensional supersymmetric field theories following
from the reduction of elliptic hypergeometric integrals to hyperbolic
$q$-hypergeometric integrals. It may be interesting to better understand
from a field theory perspective
how the various limits considered here could be realised, in particular
for the more detailed reduction scheme of \cite{BultPhD}, and
the extent to which it applies to generic CS theories with non-zero CS level.

The reduction procedure described in this paper is \textit{general} and can be
applied to any SCI for $4d$ supersymmetric theories
to give PFs for $3d$ models. Every $4d$ duality out of the large list described
in \cite{SV1,SV2}, after appropriate reduction, yields a $3d$ analogue
of the Seiberg duality similar to \cite{Aharony:1997bx,K,A}.
The  approach here provides a powerful tool
for indicating new $3d$ dualities. The absence of efficient physical checks of $3d$
dualities, such as 't Hooft anomaly matching conditions, which are useful
in $4d$,  lends added significance to this approach.
 To understand the whole tree of reductions
to PFs for $3d$ field theories one should investigate the degeneration of
elliptic hypergeometric integrals to the $q$-hypergeometric ones using the
rigorous procedure of  \cite{rai:limits}. For example,
the reduction of SCIs for $4d$ $\mathcal{N}=1$ SYM dual theories with
$SP(2N)$ gauge group and $2N_f$ quarks \cite{Intriligator1,DO,SV2}
leads to $3d$ dualities for SYM theories of \cite{A} and for CS theories
of \cite{Giveon:2008zn}, and a number of new examples.
The equality of PFs for these $3d$
$\mathcal{N}=2$ SYM and CS theories and their duals was explicitly
checked in \cite{Willett:2011gp} using the
results from \cite{BultPhD}. Starting from the equality for
the corresponding hyperbolic $q$-hypergeometric integrals (interpreted as
PFs) the dualities for $3d$ $\mathcal{N}=2$ CS theory based on $SP(2N)_k$ gauge group
with  fundamental matter were derived. The case
described in \cite{Jafferis:2011ns} should correspond to the
reduction of $4d$ $\mathcal{N}=1$ SYM theory with
$SP(2)$ gauge group, $2N_f$ quarks, and one chiral superfield in the
adjoint representation.

From the point of view of $3d$ PFs being obtained as limits in $4d$ SCIs, the results of $Z$-extremization for PFs for $3d$
theories \cite{Jafferis:2010un} may not be so unexpected since
the $a$-maximization of \cite{IW} seems to be related to certain
automorphic properties of elliptic hypergeometric integrals
describing $4d$ SCIs \cite{appear}. We hazard a guess that the
reduction to hyperbolic $q$-hypergeometric integrals preserves
some of the needed automorphic properties leading exactly to $Z$-extremization.

As shown in \cite{S4}, elliptic hypergeometric integrals emerge in
the context of Calogero-Sutherland type models either as wave
functions or as normalization conditions of wave functions.
It was conjectured there that all elliptic beta integrals
and their higher order extensions should be associated with
quantum integrable systems of such kind. Naturally, using
the connection between SCIs of four-dimensional field theories
and elliptic hypergeometric integrals, in \cite{SV2}
it was proposed that SCIs themselves have an added interpretation
in terms of relativistic Calogero-Sutherland models. The hyperbolic
$q$-hypergeometric integrals obtained in our paper (i.e., $3d$
PFs obtained by reduction of $4d$ indices)
should play a similar role in solvable hyperbolic Calogero-Sutherland
models. Also, as discussed in \cite{Spiridonov:2010em},
the hyperbolic $q$-hypergeometric integrals describe the
star-triangle relations in some continuous spin models,
i.e. they emerge as statistical sums of some two-dimensional
integrable systems.

A recent intriguing conjecture made in
\cite{Hosomichi:2010vh} concerns a connection of the partition function
for $3d$ $\mathcal{N}=4$ SYM theory with $SU(2)$ gauge group with a
kernel of the $2d$ Liouville field theory connecting conformal
blocks in different channels \cite{Teschner:2003at} (see also
\cite{Ponsot:2001ng}). This observation deserves further detailed
investigation since from our perspective these kernels can be
obtained by an appropriate reduction of the elliptic hypergeometric
integrals pushing the $4d/3d$ correspondences of the present work
down to a new $4d/2d$ correspondence.

\smallskip

\noindent {\bf Acknowledgments.} The authors are indebted to I.
Bandos, Z. Komargodski, I. Melnikov, S. Theisen,  and K. Zarembo for
stimulating discussions. G.V. would like to thank H. Nicolai for
general support. The anonymous referee is thanked for many suggestions for
improving the paper.
F.A.D. and V.S. thank the Albert-Einstein-Institut in
Golm for its warm hospitality during their visits at different
times. F.A.D. and G.V. would like to thank the organizers of the
``Simons Workshop on Mathematics and Physics $2010$" at Stony Brook
for providing a fruitful atmosphere. This work was partially
supported by RFBR (grant no.~09-01-00271) and the Heisenberg-Landau
program.

\end{document}